\documentclass[preprint,prl,showpacs]{revtex4}%
\usepackage{amsfonts}
\usepackage{amsmath}
\usepackage{amssymb}
\usepackage{graphicx}%
\begin{document}
\title{From microscopic dynamics to macroscopic irreversibility }
\author{A. P\'{e}rez-Madrid}
\affiliation{Departament de F\'{\i}sica Fonamental , Facultat de F\'{\i}sica, Universitat
de Barcelona. Diagonal 647, 08028 Barcelona, Spain}

\begin{abstract}
In this contribution we prove that the entropy of an N-body isolated system
can not decrease and the entropy production should be non-negative provided
the system possesses an equilibrium state. We define the entropy as a
functional of the set of n-particle reduced density operators ($n\leq N$)
generalizing the von Neumann fine-grained entropy formula. Additionally, as a
consequence of our analysis we find the expression of the equilibrium
n-particle reduced density operators which enter the definition of the entropy
as well as the dissipated energy in an irreversible process.

\end{abstract}
\maketitle

\textit{Introduction}.- It is a widely recognized fact that a general
mathematical theoretical proof of the second law is still lacking. As stated
in Ref. \cite{cohen} and quoted here just for illustration sake "To the best
of our knowledge no theoretical mathematical derivation of the second law has
been given up until now; instead it has been based on Kelvin's or Clausius's
principles of the impossibility of perpetual motion of the second kind
\cite{kampen}, which are based on experiment\cite{crooks}". This lack of
definitive theoretical proof has lead to reports on the violation of the
second law \cite{evans} or tests over its validity in some particular
cases\cite{abe}, \cite{ford}.

The first significant contribution to the interpretation of the second law of
Thermodynamics and the explanation of irreversibility goes back to Boltzmann.
Nevertheless, it is known that Boltzmann's contribution was criticized by
arguing that this contradicts the predictions based on the microscopic
equations of motion. Later, Gibbs and P. Ehrenfest \& T. Ehrenfest worked on
this problem by introducing coarse-graining. However, those coarse-graining
analyses require the introduction of a \textit{priori }equal probability
principles, which are hard to justify on physical grounds as was criticized by
Einstein\cite{cohen2}.

In this scenario, generalizing the Gibbs-von Neumann's statistics, it is our
contention to provide a mathematical theoretical proof of the second law for
an adiabatically isolated system based on first principles without
coarse-graining. Our starting point is the description of the state of an
isolated N-body system in terms of the set of n-particle reduced density
operators in the framework of the BBGKY [Bogolyubov-Born-Green-Kirkwood-Yvon]
description \cite{bogoliubov}. Notice that a global description of the system
in terms of the full N-particle density operator is justified for equilibrium
systems. Nonetheless, the nonuniformity of nonequilibrium systems leads to a
random clusterization and the same global quantity may be obtained for\ an
infinite number of realizations. This fact is taken into account in the BBGKY
hierarchy making this an appropriate framework for the description of
nonequilibrium systems\cite{martynov}. In this context, since the collisions
are made explicit through the collision term in the equations of motion, the
n-particle reduced density operators are not constants of motion, therefore a
way of defining the entropy to embody irreversibility might be in terms of
\ this set of reduced density operators. This is what we do here: we propose a
functional of the set of n-particle reduced density operators which
generalizes the von Neumann relative entropy as the nonequilibrium entropy of
the isolated N-body system. We will show that this entropy can not decrease
and its rate of change or entropy production, should be non-negative. In
addition, as a consequence of obtaining the entropy production we find the
expression of the equilibrium n-particle reduced density operators. In other
words, we can know nothing about the equilibrium state without previously
solving the dynamics of the system, an idea which coincides with Einstein's
point of view on the subject\cite{cohen2}.

\textit{Hamiltonian Dynamics.- }Let's consider a dynamical system of N
identical particles whose Hamiltonian is given by the N-particle Hamiltonian%
\begin{equation}
\mathbf{H}^{(N)}=\sum_{j=1}^{N}\mathbf{H}(j)+\frac{1}{2}\sum_{j\neq k=1}%
^{N}\mathbf{\phi}\left(  j,k\right)  \text{ ,} \label{hamiltonian}%
\end{equation}
where $\mathbf{H}(j)$ denotes the individual energy of the $j$-th particle and
$\mathbf{\phi}\left(  j,k\right)  $ the interaction energy between the $j$-th
and $k$-th particles. The state of the system is completely specified at a
given time by the N-particle density operator $\mathbf{D}^{(N)}(1,...,N)$
which evolves according to the Liouville-von Neumann equation
\cite{bogoliubov}
\begin{equation}
i\hbar\frac{\partial}{\partial t}D_{\alpha\beta}^{(N)}=L_{\alpha\beta
,\gamma\delta}D_{\gamma\delta}^{(N)}\text{ ,} \label{liouville}%
\end{equation}
where $D_{\alpha\beta}^{(N)}=\left\langle \alpha\right\vert \mathbf{D}%
^{(N)}\left\vert \beta\right\rangle $ are matrix elements and the Liouvillian
operator $\mathbf{L}$ \ is defined through
\begin{equation}
L_{\alpha\beta,\gamma\delta}=\left(  H_{\alpha\delta}^{(N)}\delta_{\beta
\gamma}-H_{\gamma\beta}^{(N)}\delta_{\alpha\delta}\right)  \text{ \ .}
\label{liouvillian}%
\end{equation}
Here, repeated Greek indices mean summation. However, a strictly equivalent
alternative description of the state of the system can be given in terms of
the set of n-particle reduced density operators \cite{bogoliubov},
\cite{dufty}
\begin{equation}
\mathcal{D}\equiv\left\{  \mathbf{D}^{(0)},\mathbf{D}^{(1)},......,\mathbf{D}%
^{(N)}\right\}  \text{ \ ,} \label{superoperator}%
\end{equation}
where $\mathcal{D}$ constitutes a density superoperator whose components are
the n-particle reduced density operators $\mathbf{D}^{(n)}(1,....,n)$ obtained
by taking the trace over $N-n$ particles, namely,
\begin{equation}
\mathbf{D}^{(n)}(1,....,n)=\frac{N!}{(N-n)!}\underset{(n+1,...,N)}{\text{Tr}%
}\mathbf{D}^{(N)}(1,....,N)\text{ \ ,} \label{reduced density}%
\end{equation}
and $D_{\alpha\beta}^{(0)}=\delta_{\alpha\beta}$ (\textit{i.e.} $\mathbf{D}%
^{(0)}$ coincides with the unit matrix). These operators satisfy the
normalization condition
\begin{equation}
\underset{_{\left(  1,...,n\right)  }}{\text{Tr}}\left\{  \mathbf{D}%
^{(n)}(1,....,n)\right\}  \text{ =}\frac{N!}{(N-n)!}\text{ \ .}
\label{normalization}%
\end{equation}
On the other hand, the dynamics of the $\mathbf{D}^{(n)}$ follows after
performing the operation $\underset{(n+1,...,N)}{\text{Tr}}$ on both sides of
Eq. (\ref{liouville}). Thus, we obtain the so-called quantum BBGKY hierarchy
of equations
\begin{align}
&  i\hbar\frac{\partial}{\partial t}D_{\alpha\beta}^{(n)}-\left(
H_{\alpha\delta}^{(n)}D_{\alpha\gamma}^{(n)}-H_{\gamma\beta}^{(n)}%
D_{\delta\beta}^{(n)}\right) \nonumber\\
&  =\underset{(n+1)}{\text{Tr}}\left\{  F_{\alpha\delta}^{\left(  n+1\right)
}D_{\alpha\gamma}^{(n+1)}-F_{\gamma\beta}^{(n+1)}D_{\delta\beta}%
^{(n+1)}\right\}  \text{ \ ,} \label{hierarchy}%
\end{align}
where $\mathbf{H}^{(n)}$ is the n-particle Hamiltonian
\begin{equation}
\mathbf{H}^{(n)}=\sum_{j=1}^{n}\mathbf{H}(j)+\frac{1}{2}\sum_{j\neq k=1}%
^{n}\mathbf{\phi}\left(  j,k\right)  \text{ } \label{nhamiltonian}%
\end{equation}
and
\begin{equation}
\mathbf{F}^{\left(  p\right)  }(1,....,p)=%
{\displaystyle\sum\limits_{j=1}^{p-1}}
\mathbf{\phi}\left(  j,p\right)  \text{ \ .} \label{interaction}%
\end{equation}
The second part on the left hand side of Eq. (\ref{hierarchy}) constitutes the
streaming term which gives the unitary evolution while the contribution on the
right hand side is the collision term responsible for the non-unitary
evolution of $\mathbf{D}^{(n)}$. Streaming and collision terms balance out at
equilibrium when $\mathbf{D}^{(n)}=\mathbf{D}_{eq}^{(n)}$.

\textit{Entropy and irreversibility.- }In view of the fact that the
statistical state of the system is given in terms of the set of reduced
density operators, we define the nonequilibrium entropy
\begin{align}
S  &  =-k_{B}\text{Tr}\left\{  \mathcal{D}\left(  \ln\mathcal{D-}%
\ln\mathcal{D}_{eq}\right)  \right\}  +S_{eq}\nonumber\\
&  =-k_{B}\sum_{n=0}^{N}\frac{1}{n!}\text{Tr}\left\{  \mathbf{D}^{(n)}\left(
\ln\mathbf{D}^{(n)}-\ln\mathbf{D}_{eq}^{(n)}\right)  \right\}  \;+S_{eq}\text{
,} \label{gibbs}%
\end{align}
a functional of the density superoperator $\mathcal{D}$ which constitutes a
generalization of the relative or conditional von Neumann entropy. This
entropy is the sum of all the n-particle entropies, a definition which is not
redundant since the nonequilibrium system behaves as a random mixture of
n-particle systems. Here $S_{eq}$ is the equilibrium entropy which corresponds
to the equilibrium density superoperator $\mathcal{D}_{eq}$ (=$\left\{
\mathbf{D}^{(0)},\mathbf{D}_{eq}^{(1)},......,\mathbf{D}_{eq}^{(N)}\right\}
$). This equilibrium density superoperator is obtined when the right hand side
Eq. (\ref{hierarchy}) is balanced by the second term on the left hand side of
this equation, which occurs in a time scale related to the
collisions\cite{huang}. Moreover, $S_{eq}$ must coincides with the
thermodynamic entropy given by the Clausius's, Boltzmann's, Gibbs's etc. entropies.

It should be pointed out that in defining the nonequilibrium entropy through
Eq. (\ref{gibbs}), we assume that the BBGKY hierarchy describes a mixture of
`compressible' fluids, each constituted by the set of n-particle clusters of
the same size n. The origin of this `compressibility' is the interaction
between different fluids which leads to the creation of n-particles clusters
at the expense of the annihilation of p-particle clusters with $n\neq p$. Each
fluid contributes with its entropy, the n-particle entropy, to the total
nonequilibrium entropy of our system. Moreover, an additional consequence of
the molecular collisions is the fact that each of the n-particle entropies can
fluctuate and therefore also $S$. In the thermodynamic limit the fluctuations
of $S$ mentioned here should be gaussian, nevertheless this is not necessarily
the case for finite $N$.

More interestingly here the most important property of this entropy is its
direction of change in a natural process. To elucidate this, we first note
that since $\mathbf{X}\ln\mathbf{X}$ is a convex function and $\mathbf{D}%
^{(n)}$ and $\mathbf{D}_{eq}^{(n)}$ both are density operators \cite{balian}%
\begin{equation}
\text{Tr}\left\{  \mathbf{D}^{(n)}\left(  \ln\mathbf{D}^{(n)}-\ln
\mathbf{D}_{eq}^{(n)}\right)  \right\}  \geq0 \label{inequality}%
\end{equation}
which given $\mathbf{D}^{(n)}=\sum_{m}D_{m}^{(n)}\left\vert m\right\rangle
^{(n)(n)}\left\langle m\right\vert $ and $\mathbf{D}_{eq}^{(n)}=\sum
_{r}D_{eq,r}^{(n)}\left\vert r\right\rangle ^{(n)(n)}\left\langle r\right\vert
$ where $D_{m}^{(n)}$ and $D_{eq,r}^{(n)}$ are eigenvalues and $\left\vert
m\right\rangle ^{(n)}$ and $\left\vert r\right\rangle ^{(n)}$ eigenvectors,
can be written%
\begin{equation}
\sum_{q,m}\left\vert ^{(n)}\langle r\mid m\rangle^{(n)}\right\vert
^{2}\left\{  D_{m}^{(n)}\left(  \ln D_{m}^{(n)}-\ln D_{eq,r}^{(n)}\right)
\right\}  \geq0\text{ .} \label{inequality_2}%
\end{equation}
Additionally since $D_{m}^{(n)}$ gives the probability of the state
$\left\vert m\right\rangle $ at time $t$, it should be bounded so that%
\begin{equation}
0\leq\sum_{q,m}\left\vert ^{(n)}\langle r\mid m\rangle^{(n)}\right\vert
^{2}\left\{  D_{m}^{(n)}\left(  \ln D_{m}^{(n)}-\ln D_{eq,r}^{(n)}\right)
\right\}  \leq C \label{inequality_3}%
\end{equation}
for a given constant $C$, which implies that the entropy $S$ possesses a lower
bound%
\begin{equation}
S-S_{eq}=-k_{B}\sum_{n=0}^{N}\frac{1}{n!}\sum_{q,m}\left\vert ^{(n)}\langle
r\mid m\rangle^{(n)}\right\vert ^{2}\left\{  D_{m}^{(n)}\left(  \ln
D_{m}^{(n)}-\ln D_{eq,r}^{(n)}\right)  \right\}  \geq-k_{B}\sum_{n=0}^{N}%
\frac{1}{n!}C\text{ \ .} \label{bound}%
\end{equation}
Thus, one concludes that the nonequilibrium entropy can not decrease.

\textit{Entropy production.- }Following the\textit{ }previous line of thought,
by using the BBGKY hierarchy Eq. (\ref{hierarchy}) one can compute the rate of
change of $S$ Eq. (\ref{gibbs}), \textit{i.e.} the entropy production,
obtaining%
\begin{align}
\frac{\partial}{\partial t}S  &  =-i\hbar^{-1}k_{B}\sum_{n=0}^{N}\frac{1}%
{n!}\text{Tr}\left\{  \left[  \mathbf{H}^{\left(  n\right)  }+\mathcal{F}%
^{(n)},\mathbf{D}^{(n)}\right]  \left(  \ln\mathbf{D}_{eq}^{(n)}-\ln
\mathbf{D}^{(n)}\right)  \right\}  =\nonumber\\
&  -i\hbar^{-1}k_{B}\sum_{n=0}^{N}\frac{1}{n!}\text{Tr}\left\{  \left[
\left(  \mathbf{H}^{\left(  n\right)  }+\mathcal{F}^{(n)}\right)
,\ln\mathbf{D}_{eq}^{(n)}\right]  \mathbf{D}^{(n)}\right\}  \geq0
\label{entropy_prod}%
\end{align}
where the cyclic invariance property of the trace has been used and $\left[
..,..\right]  $ is the commutator. Moreover, $\mathcal{F}^{(n)}$ has been
defined through $\mathbf{D}^{(n)}\mathcal{F}^{(n)}\mathcal{=}\underset
{(n+1)}{\text{Tr}}\left\{  D^{(n+1)}F^{\left(  n+1\right)  },\right\}  $ does
accounting for the interaction among the fixed particles of the $n-th$ cluster
with the remaining $N-n$ particles of the system. The entropy production
(\ref{entropy_prod}) is zero at equilibrium when $\mathbf{D}^{(n)}%
=\mathbf{D}_{eq}^{(n)}$ whereas in any other case it is generally non-zero, so
in view of our conclusion after Eq. (\ref{bound}) the entropy production
should be positive. Thus, it might be said that this constitutes a general
microscopic proof of the second law.

In addition, as a consequence of Eq. (\ref{entropy_prod}), one obtains that
\begin{equation}
\left[  \mathbf{H}^{\left(  n\right)  }+\mathcal{F}_{eq}^{(n)},\ln
\mathbf{D}_{eq}^{(n)}\right]  =0 \label{equiv_condi}%
\end{equation}
which suffices to satisfy the extremum condition $\delta(\partial S/\partial
t)/\delta\mathbf{D}^{(n)}\mid_{eq}=0$ meaning that
\begin{equation}
\ln\mathbf{D}_{eq}^{(n)}=K\left(  \mathbf{H}^{\left(  n\right)  }%
+\mathcal{F}_{eq}^{(n)}\right)  \label{equiv_den}%
\end{equation}
for a given proportionality constant $K$, which after substitution in Eq.
(\ref{entropy_prod}) allows us to rewrite this equation
\begin{equation}
\frac{\partial}{\partial t}S=-i\hbar^{-1}k_{B}K\sum_{n=0}^{N}\frac{1}%
{n!}\text{Tr}\left\{  \left[  \mathcal{F}^{(n)},\mathcal{F}_{eq}^{(n)}\right]
\mathbf{D}^{(n)}\right\}  \text{ ,} \label{entropy_prod_2}%
\end{equation}
a compact way of writing $\partial S/\partial t$. Our result given through
Eqs. (\ref{entropy_prod}) and (\ref{entropy_prod_2}) is exact, thus possessing
the generality that is lacking in the Boltzmann's H-theorem restricted to the
molecular chaos regime.

In the case that the system is in contact with a heat bath one can say that
$\mathcal{D}_{eq}(t)$ and $S_{eq}(t)$, thus we can write%
\begin{equation}
\frac{\partial}{\partial t}S=\dot{\sigma}+\frac{\dot{Q}}{T_{o}}\text{
,}\label{second_law}%
\end{equation}
where $\dot{\sigma}$ is the irreversible entropy production given by right
hand side of Eq. (\ref{entropy_prod_2}) and $\dot{Q}/T_{o}=\partial
S_{eq}/\partial t$, with $\dot{Q}$ being the the rate of heat interchange and
$T_{o}$ is the bath temperature. According to our previous discussion
$\dot{\sigma}\geq0$, therefore%
\begin{equation}
\delta S\geq\frac{\delta Q}{T_{o}}\label{second_law_2}%
\end{equation}
which corresponds to the Clausius' form of the second law. Hence,%
\begin{equation}
T_{o}\delta\sigma=T_{o}\delta S-\delta Q\label{second_law_4}%
\end{equation}
coincides with the energy dissipated in the irreversible process. On the other
hand, for reversible processes $\mathcal{D}(t)=\mathcal{D}_{eq}(t)$ at every
moment, therefore $\sigma=0$ and Eq. (\ref{second_law_2}) becomes
\begin{equation}
\delta S=\frac{\delta Q}{T_{o}}\label{second_law_3}%
\end{equation}

\textit{Conclusions}.- We find that the description of the N-body system in
the framework of the BBGKY hierarchy enables us to prove that the
nonequilibrium entropy increases in a natural process, which constitutes a
general microscopic proof of the second law. We emphasize that the entropy
should be defined as a functional of the density superoperator. Likewise, our
theoretical analysis allows us to obtain the expression of the set of
equilibrium n-particle density operators, \textit{i.e.} the equilibrium
density superoperator. This fact shows that the state of equilibrium is
determained as a consequence of the dynamics and not given a \textit{priori}.

As to how irreversibility can be derived from the reversible equations of
mechanics, although the BBGKY hierarchy (\ref{hierarchy}) expressing the
microscopic dynamics is invariant under the time reversal operation, thus
representing a set of reversible equations, this fact is not contradicted by
the law of increase of nonequilibrium entropy we have derived. Let me explain,
according to Eq. (\ref{hierarchy}), a stationary state is reached when the
right hand side of this equation is balanced by the second term on the left
hand side, this being the equilibrium state since this is the only stationary
state for an isolated system\cite{tolman}. This stationarity or equilibrium
condition is the same both for positive and negative time without distinction
since we could say that at equilibrium there is no distinction between
positive and negative time because at equilibrium time does not exist. Thus,
the evolution along the time line either in the direct or inverse sense always
ends in one and the same equilibrium state. Additionally, we have shown in the
discussion after Eq. (\ref{gibbs}) that the process of relaxation towards
equilibrium entails the production of entropy according to Eqs.
(\ref{entropy_prod}) and (\ref{entropy_prod_2}), thus having the character of
an irreversible process. In conclusion, it could be said that time inversion
is a mathematical artifice since there is only one Universe whose state given
through the density superoperator (distribution functions in the classical
language) evolves in one direction, the direction along which entropy
increases up to its equilibrium value $S\_eq$, wether this direction be direct
or inverse. According to the time reversal invariance this is what one would
predict. In fact, let us suppose that we have two identical isolated systems A
and B, which at $t=0$ have positive and negative velocities, respectively . In
those conditions, the past of B is the future of A or conversely, the future
of B is the past of A. This can be reworded in the following terms, if at a
given time let us say $t=0$, an isolated system is out of the equilibrium
state, then $S$ is at a local minimum and thus both $\partial S/\partial
t\geq0$ for $t>0$ and $\partial S/\partial t\leq0$ for $t<0$, hence it follows
that $\partial S/\partial t$ is not necessarily a continuous function of time.
Therefore, in those terms there is no reversible paradox.

It should be stressed that in the basis of our treatment there is the
assumption that the state of the system at a given time is determined by an
statistical operator which is a function of the wave function, and therefore
the Poincar\'{e} recurrence theorem does not apply. In the classical language
we would say that we do not deal with trajectories in the phase space but
rather with distribution functions which determine the state of the system at
a given time. A more fundamental proof of the fact that our theory does not
incur in the recurrence paradox comes from the behavior of the entropy
production $\partial S/\partial t$ as a function of the time explained above.

To conclude, we have shown here that the reversible microscopic dynamics
induces macroscopic irreversibility.

\end{document}